\documentstyle[12pt]{article}

\catcode`\@=11
\long\def\@makefntext#1{
\protect\noindent \hbox to 3.2pt {\hskip-.9pt
$^{{\ninerm\@thefnmark}}$\hfil}#1\hfill}        

 \def\@makefnmark{\hbox to 0pt{$^{\@thefnmark}$\hss}}  

\def\ps@myheadings{\let\@mkboth\@gobbletwo
\def\@oddhead{\hbox{}
\rightmark\hfil\ninerm\thepage}
\def\@oddfoot{}\def\@evenhead{\ninerm\thepage\hfil
\leftmark\hbox{}}\def\@evenfoot{}
\def\sectionmark##1{}\def\subsectionmark##1{}}

\newcounter{sectionc}\newcounter{subsectionc}\newcounter{subsubsectionc}
\renewcommand{\section}[1] {\vspace{0.6cm}\addtocounter{sectionc}{1}
\setcounter{subsectionc}{0}\setcounter{subsubsectionc}{0}\noindent
      {\bf\thesectionc. #1}\par\vspace{0.4cm}}
\renewcommand{\subsection}[1] {\vspace{0.6cm}\addtocounter{subsectionc}{1}
      \setcounter{subsubsectionc}{0}\noindent
      {\it\thesectionc.\thesubsectionc. #1}\par\vspace{0.4cm}}
\renewcommand{\subsubsection}[1]
   {\vspace{0.6cm}\addtocounter{subsubsectionc}{1} \noindent
{\rm\thesectionc.\thesubsectionc.\thesubsubsectionc.  #1}\par\vspace{0.4cm}}

\newcounter{appendixc}
\newcounter{subappendixc}[appendixc]
\newcounter{subsubappendixc}[subappendixc]

\renewcommand{\appendix}[1] {\vspace{0.6cm}
        \refstepcounter{appendixc}
        \setcounter{figure}{0}
        \setcounter{table}{0}
        \setcounter{equation}{0}
        \renewcommand{\thefigure}{\Alph{appendixc}.\arabic{figure}}
        \renewcommand{\thetable}{\Alph{appendixc}.\arabic{table}}
        \renewcommand{\theappendixc}{\Alph{appendixc}}
        \renewcommand{\theequation}{\Alph{appendixc}.\arabic{equation}}
        \noindent{\bf Appendix \theappendixc #1}\par\vspace{0.4cm}}

\def\abstracts#1{{
      \centering{\begin{minipage}{30pc}\tenrm\baselineskip=12pt\noindent
      \centerline{\tenrm ABSTRACT}\vspace{0.3cm}
      \parindent=0pt #1
      \end{minipage}}\par}}

\renewenvironment{thebibliography}[1]
      {\begin{list}{\arabic{enumi}.}
      {\usecounter{enumi}\setlength{\parsep}{0pt}
\setlength{\leftmargin 1.25cm}{\rightmargin 0pt}
       \setlength{\itemsep}{0pt} \settowidth
      {\labelwidth}{#1.}\sloppy}}{\end{list}}

\newcounter{itemlistc}
\newcounter{romanlistc}
\newcounter{alphlistc}
\newcounter{arabiclistc}

\newcommand{\fcaption}[1]{
        \refstepcounter{figure}
        \setbox\@tempboxa = \hbox{\tenrm Fig.~\thefigure. #1}
        \ifdim \wd\@tempboxa > 6in
           {\begin{center}
        \parbox{6in}{\tenrm\baselineskip=12pt Fig.~\thefigure. #1}
            \end{center}}
        \else
             {\begin{center}
             {\tenrm Fig.~\thefigure. #1}
              \end{center}}
        \fi}

\newcommand{\tcaption}[1]{
        \refstepcounter{table}
        \setbox\@tempboxa = \hbox{\tenrm Table~\thetable. #1}
        \ifdim \wd\@tempboxa > 6in
           {\begin{center}
        \parbox{6in}{\tenrm\baselineskip=12pt Table~\thetable. #1}
            \end{center}}
        \else
             {\begin{center}
             {\tenrm Table~\thetable. #1}
              \end{center}}
        \fi}

\def\@citex[#1]#2{\if@filesw\immediate\write\@auxout
      {\string\citation{#2}}\fi
\def\@citea{}\@cite{\@for\@citeb:=#2\do
      {\@citea\def\@citea{,}\@ifundefined
      {b@\@citeb}{{\bf ?}\@warning
      {Citation `\@citeb' on page \thepage \space undefined}}
      {\csname b@\@citeb\endcsname}}}{#1}}

\newif\if@cghi
\def\cite{\@cghitrue\@ifnextchar [{\@tempswatrue
      \@citex}{\@tempswafalse\@citex[]}}
\def\citelow{\@cghifalse\@ifnextchar [{\@tempswatrue
      \@citex}{\@tempswafalse\@citex[]}}
\def\@cite#1#2{{$\null^{#1}$\if@tempswa\typeout
      {IJCGA warning: optional citation argument
      ignored: `#2'} \fi}}

\def\fnt#1#2{\footnotetext{\kern-.3em
      {$^{\mbox{\sevenrm #1}}$}{#2}}}

 1
 1
 1

\font\tenbf=cmbx10
\font\tenrm=cmr10
\font\tenit=cmti10

\font\ninerm=cmr9

\let\a=\alpha 
\let\g=\gamma \let\d=\delta

\let\m=\mu

\def\2{{1\over2}} \def\4{{1\over4}} \def\52{{5\over2}}
\def\6{\partial }

\def\({\left(} \def\){\right)} \def\<{\langle } \def\>{\rangle }

\def\CO{{\cal O}}

\def\beg{\begin{equation}}
\def\begar{\begin{eqnarray}}
\def\ee{\end{equation}}
\def\ea{\end{eqnarray}}

\begin{document}

\hfill TUW-94-15

\vspace{1cm}

\centerline{\tenbf WEAK COUPLING LIMIT AND GENUINE QCD
PREDICTIONS}
\baselineskip=22pt
\centerline{\tenbf FOR HEAVY QUARKONIA\footnote{Presented at the Int.\
Conference \lq\lq Quark Confinement and the Hadron Spectrum", Villa Olmo-Como,
Italy, June 20-24, 1994} }
\vspace{0.8cm}
\centerline{\tenrm WOLFGANG KUMMER}
\centerline{\tenrm and}
\centerline{\tenrm WOLFGANG M\"ODRITSCH}
\baselineskip=13pt
\vspace{0.3cm}
\centerline{\tenit Institut f\"ur Theoretische Physik, Technische
Universit\"at Wien}
\baselineskip=12pt
\centerline{\tenit Wiedner Hauptstra\ss e 8-10, A-1040 Wien,
Austria }
\vspace{0.9cm}
\abstracts{Although individual levels of toponium will be unobservable,
the top--anti--top system near threshold fulfills all requirements of a
rigorous perturbation theory in QCD for weakly bound systems.
Corresponding techniques from positronium may thus be transferred
successfully to this case. After clarifying the effect of a non-zero
width we calculate the $t\bar{t}$ potential to be used for the
calculation of e.g. the cross-sections near threshold.}

\vspace{0.8cm}
\rm\baselineskip=14pt


Perturbative quantum field theory by means of the general
Bethe--Salpeter (BS) formalism is applicable also for weakly bound
systems, described as corrections to a zero order equation for the
bound system. The latter is usually taken as the
Schr\"odinger--equation, but experience in positronium has shown that
other starting points of perturbation theory, like the Barbieri-Remiddi
(BR) equation,\cite{bar78} offer advantages.
That such rigorous methods \cite{lep77} have not found serious considerations
for a long time in quarkonia has to do with the importance of confinement
effects in QCD. Parametrizing the latter by a gluon-condensate \cite{Shif},
one finds that for quark masses $m$ up to about 50 GeV nonperturbative effects
are of the same order as corrections to the bound states \cite{Leyt}, athough
the situation seems a little less serious using
BS-methods\nolinebreak\cite{WKGW}.  Nevertheless, the possible practical
advantages of a rigorous approach have been demonstrated a long time ago, not
for level shifts in bound-states, but for perturbative corrections to the
decay of quark-antiquark systems.  Straightforward applications of
perturbative QCD to the annihilation part {\it alone} of, say,
bottom-anti-bottom, with minimal subtraction at a scale $O(2m)$ gave huge a
correction $O(10 \alpha_s/\pi)$\nolinebreak \cite{Barb}.  However, in such a
decay the bound quark pair is really off-shell. Although from that it is clear
that an (off-shell) annihilation part by itself cannot even be gauge
independent: The perturbative correction to the wave-function of the decaying
state must be added for consistency. In fact, after doing this and performing
a careful renormalization procedure at Bohr momentum $O(\alpha_s m)$,
appropriate for that system, it could be shown that large corrections tend to
compensate in that result, at least in the case of the singlet ground state
($0^{-+}$) \cite{KumW}.

Now, with a top quark in the range of 150 - 180 GeV \cite{lepcdf} for the
first time confinement effects are negligible. The large width $\Gamma$ of the
decay $t \to b + W$ makes bound--state 'poles' at real energies unobservable,
but at the same time obliterates confinement effects even {\it above}
threshold because the top quark has no time to 'hadronize'.  The effect of
$\Gamma$ can be taken into account by simple analytic continuation of the zero
order equation to complex energies \cite{fad88}. Previous work \cite{sum93}
was based upon phenomenological quarkonium potentials. However also BS
perturbation theory can be applied to the $t\bar t$ Green function G near the
poles which move into the unphysical sheet of the complex plane at a distance
$\Gamma$ to the real energy axis. Nevertheless, the residues and hence the
wave functions together with the QCD level shifts remain real. Hence they can
be used in a straightforward manner to determine the different contributions
to a rigorous QCD potential.

After showing that this analytic continuation argument also works for the
BR-equation, we determine the different contributions for such a quantity from
the (real) energy shifts, including (numerical) $\CO (m\alpha^4_s)$--effects.
With a generic perturbation H to the Green function $G_0$ of the zero order
equation, the level shifts become \cite{lep77}
\begin{equation}
\Delta E_n = \langle\langle h_i \rangle\rangle (1 + \langle\langle
h_1\rangle\rangle ) + \langle\langle h_0 g_1 h_2 \rangle\rangle +
\CO(h^3)
\end{equation}
where $h_i$ and $g_i$ are the expansion coefficients of H, resp. $G_0$ near
the pole $E \sim E_n$ of $G_0$.  The expectation values are {\it
four}--dimensional momentum integrals, taken here with respect to BR
wavefunctions. The latter differ by factors produced by relativistic
corrections and with $p_0$ from (normalized) Schr\"odinger--wavefunctions.
Nevertheless, we formulate our final result as a 'potential'. The full formula
for the different parts of
\begin{equation}
V = \sum_{i=0}^2 V_{QCD}^{(i)} + V_{EW}
\end{equation}
can be found in ref.\cite{kum94}. $(i)$ refers to the loop order which, of
course, is not uniquely correlated with orders of $\alpha_s$ in energy shifts.
--- $V^{(0)}_{QCD} $ beside the Coulomb exchange contains the
$\vec{p}\,^4$--term and the exchange of a transversal gluon, producing the
'abelian' relativistic corrections $\CO (m\alpha_s^4)$. We {\it do not}
include a running coupling constant anywhere because this would mix orders,
spoiling even the gauge--independence order by order within any application.
Of course, for technical reasons eq.(2) is derived and should be applied in
the Coulomb gauge.  $$V^{(1)}_{QCD} =- \frac{33 \a^2}{8 \pi r} ( \g + \ln \m
r) +\frac{\a^2}{4\pi r} \sum_{j=1}^{5} [\mbox{Ei}(-r m_j e^{\frac{5}{6}})
-\frac{5}{6} + \2 \ln (\frac{\m^2}{m_j^2} + e^{\frac{5}{3}})] + \frac{9
\a^2}{8 m r^2} $$ consists of vacuum polarization and vertex corrections. The
first contain the gluon loop and the loops from fermions. Although the level
contributions will be $\CO (m\alpha_s^3)$ for the toponium system the mass of
the charm and bottom must not be neglected, because they yield {\it numerical}
$\CO (m\alpha_s^4)$ corrections. In contrast to the QED case here also the
one--loop gluon--splitting vertex yields a potential $\propto \alpha_s^2 /
mr^2$, important to this order.  In the vacuum polarization part of
\begin{eqnarray}V^{(2)}_{QCD} &=&  c^{(H)} \frac{4\pi \a^3}{r} \nonumber \\
      & &- 2 \frac{\a^3}{(16 \pi)^2 r} \left\{ (33-2n_f)^2[\frac{\pi^2}{6}+
2(\g+\ln \m r)^2] + 9 (102-\frac{38}{3} n_f)(\g+\ln \m r) \right\} \nonumber
\end{eqnarray}
we emphasize the importance of {\it non-leading} logarithms which would only
be contained in the usual running coupling constant if a two--loop
$\beta$--function would be used.  Among the 'box' corrections, an H--graph
(with the figure H formed by gluons between the two fermion lines) is
emphasized as an contribution which gives at least a correction to the Coulomb
term. In
\begin{eqnarray}
       V_{EW} &=& - \frac{8}{9} \a_{QED}(\m) \frac{4\pi \a}{r} - \sqrt{2} G_F
m^2 \frac{e^{-m_H r}}{4 \pi r} + \sqrt{2} G_F m_Z^2 a_f^2
\frac{\d(\vec{r})}{m_Z^2} (7- \frac{11}{3} \vec{S}\,^2 )  \nonumber \\ & & +
\sqrt{2} G_F m_Z^2 a_f^2  \frac{e^{-m_Z r}}{2 \pi r}
\big[1-\frac{v_f^2}{2a_f^2} - (\vec{S}\,^2 - 3 \frac{(\vec{S}\vec{r})^2}{r^2})
(\frac{1}{m_Z r} + \frac{1}{m_Z^2 r^2})- (\vec{S}\,^2 -
\frac{(\vec{S}\vec{r})^2}{r^2})\big], \nonumber
\end{eqnarray}
photon--exchange, Z--exchange and
Z--anihilation turn out to be as essential for the $t\bar t$--system
as the usual relativistic corrections \cite{kum94}.

\vspace{.5cm}
One of the authors (W.M) thanks the organizers for financial
support.This work is supported in part by the Austrian Science Foundation
(FWF) in project P10063-PHY within the framework of the EEC- Program
"Human Capital and Mobility", Network "Physics at High Energy Colliders",
contract CHRX-CT93-0357 (DG 12 COMA).

\end{document}